\documentclass[prd,showpacs]{revtex4}
\usepackage{epsfig}
\usepackage{amsmath}
\usepackage{amssymb}

\begin{document}
\title{Study of W-exchange Mode $D^0 \to \phi \overline K^0$}\thanks{This
work is partly supported by National Science Foundation of China
under contract No. 10475085}
\author{
 Dong-Sheng Du,  Ying Li\footnote{liying@mail.ihep.ac.cn},  Cai-Dian L\"u} \affiliation
 {\it \small Institute of High
Energy Physics, P.O.Box 918(4), Beijing 100049, China}
\begin{abstract}
We calculate the branching ratio of rare decay $D^0 \to \phi
\overline K^0$ using the  perturbative QCD factorization approach based on $k_T$
factorization. Our result shows this branching ratio is $(8.7
\pm1.4)\times 10^{-3}$, which is consistent with experimental
data.  We hope the CLEO-C and BES-III can measure it more
accurately, which will help us  understand QCD dynamics and
$D$ meson weak decays.
\end{abstract}

\pacs{13.20.Ft, 12.38.Bx, 14.40.Lb}
\maketitle

The precise estimation of the branching ratio for the hadronic $D$
decays is very important both in theoretical side and experimental
side. As a rare decay, $D^0 \to \phi \overline K^0$ plays an
important role in testing QCD dynamics and searching for new
physics.  In the standard model picture, this decay mode is a pure
annihilation type decay, also called $W$-exchange mode.   It has
been discussed with a large branching ratio to explain the big
difference of lifetime between $D^0$ and $D^+$ \cite{bigi}. In the
$80$'s of last century, this decay's branching ratio has been
measured in CLEO \cite{benek}. With the development of experiments,
this branching ratio is confirmed as \cite{pdg}:
\begin{gather}
 \mathrm{Br}(D^0 \to \phi \overline K^0) = (9.4 \pm{1.1})\times
 10^{-3}.\label{exp}
\end{gather}
To our knowledge, this decay mode is never calculated quantitatively
in QCD. So, it is necessary for us to reanalyze this decay mode
seriously.

Based on the factorization hypothesis, many $D$ decays have been
calculated in naive factorization approach \cite{BSW} and QCD
factorization approach \cite{QCDF}. However, Annihilation diagrams
can not be calculated easily for its endpoint singularity. It is
well known  that perturbative QCD (PQCD) factorization approach is
successful in calculating two-body $B$ meson decays \cite{pqcd}.
The end point singularity can be regulated by Sudakov form factor and
threshold resummation through introducing the transverse momentum
$k_T$ of valence quarks. Thus, PQCD approach can give converging
results and have predictive power.   Using this approach, people
have calculated many $B$ meson pure annihilation type decays quantificationally, and
the results agree with data well \cite{lu:bdsk}.

In this paper, we will calculate the $D^0 \to \phi \overline K^0$
decay in PQCD approach. In this decay, the $W$ boson exchange
induces the four quark operator $\bar s c \to \bar{u}d$, and the
$\bar{s}s$ quarks included in $\phi, \overline K^0$ are produced
 from a gluon. This gluon can attach to any one of the quarks
participating in the four-quark operator. In the rest frame of $D$
meson, the produced $s$ and $\bar{s}$ quark included in final states
have momenta of $\mathcal{O}(P_K/2)$ and $\mathcal{O}(P_\phi/2)$,
respectively.  Therefore the gluon producing them has momentum $q \sim
\mathcal{O}(M_D/2)$, which is nearly $1$ GeV. This hard gluon can be
treated perturbatively. Therefore the hard part calculation involves
not only the four quark operator but also the hard gluon connecting $s \bar s$
quark pair. The factorization here means the six quark hard part
calculation factorize from the non-perturbative hadronization
characterized by meson light cone wave functions.

We work at the  $D$ meson  rest frame for simplicity. In
light-cone coordinate system, the momentum of the $D$, $\phi$ and
$\overline K^0$ meson can be written as:
\begin{eqnarray}
P_1=P_D=\frac{M_D}{\sqrt{2}} ~(1, 1, \vec{0}_\perp),
P_2=P_{\phi}=\frac{M_D}{\sqrt{2}} ~(1, r^2, \vec{0}_\perp),
P_3=P_{\overline K^0}=\frac{M_D}{\sqrt{2}}~ (0, 1-r^2,
\vec{0}_\perp),
\end{eqnarray}
where $r = M_\phi/ M_D $ and we neglect the $\overline K^0$ meson
mass $M_{\overline K^0}$ compared with the large $D$ meson mass.
Because this decay is $D \to PV$ mode, the transverse polarization
of $\phi$ meson gives no contribution. The longitudinal polarization
vector of $\phi$ meson is given as:
\begin{eqnarray}
\epsilon_{2L}=\frac{M_D}{\sqrt{2}M_\phi} ~(1, -r^2,\vec{0}_\perp).
\end{eqnarray}
Denoting the light (anti-)quark momenta in $D$, $\phi$ and
$\overline K^{0}$ mesons as $k_1$, $k_2$, and $k_3$, respectively,
we can choose
\begin{eqnarray}
k_1 = (x_1P_1^+, 0, {\bf k}_{1T}),\ k_2 = (x_2 P_2^+, 0, {\bf
k}_{2T}), k_3 = (0, x_3 P_3^-, {\bf k}_{3T}) .
\label{eq:momentun2}
\end{eqnarray}
Then, integrating over $k_1^-$, $k_2^-$, and $k_3^+$, we  get the decay
amplitude:
\begin{multline}
 \mbox{Amplitude}
\sim \int\!\! d x_1 d x_2 d x_3
b_1 d b_1 b_2 d b_2 b_3 d b_3 \\
\mathrm{Tr} \bigl[ C(t) \Phi_D(x_1,b_1)
\Phi_{\phi}(x_2,b_2,\epsilon) \Phi_{K}(x_3, b_3)
 H(x_i,b_i,\epsilon, t) S_t(x_i)\, e^{-S(t)} \bigr],
\label{eq:convolution2}
\end{multline}
where $b_i$ is the conjugate space coordinate of ${\bf k}_{iT}$, and $t$
is the largest energy scale in $H$, as the function in terms of
$x_i$ and $b_i$. The large logarithms ($\ln m_W/t$) coming from
QCD radiative corrections to four quark operators are included in
the Wilson coefficients $C(t)$. The large double logarithms
($\ln^2 x_i$) on the longitudinal direction are summed by the
threshold resummation, and they lead to a jet function $S_t(x_i)$
which smears the end-point singularities on $x_i$. The last term,
$e^{-S(t)}$, contains two kinds of logarithms. One of the large
logarithms is due to the renormalization of ultra-violet
divergence $\ln tb$, the other is resummation of double logarithm $\ln^2 b$ from
the overlap of collinear and soft gluon corrections. This Sudakov
form factor suppresses the soft dynamics effectively. Thus it
makes perturbative calculation of the hard part $H$ reliable.
$\Phi_M$ is the wave function which describes the inner
information of meson $M$.

\begin{figure}[htb]
\begin{center}
\includegraphics[scale=0.45]{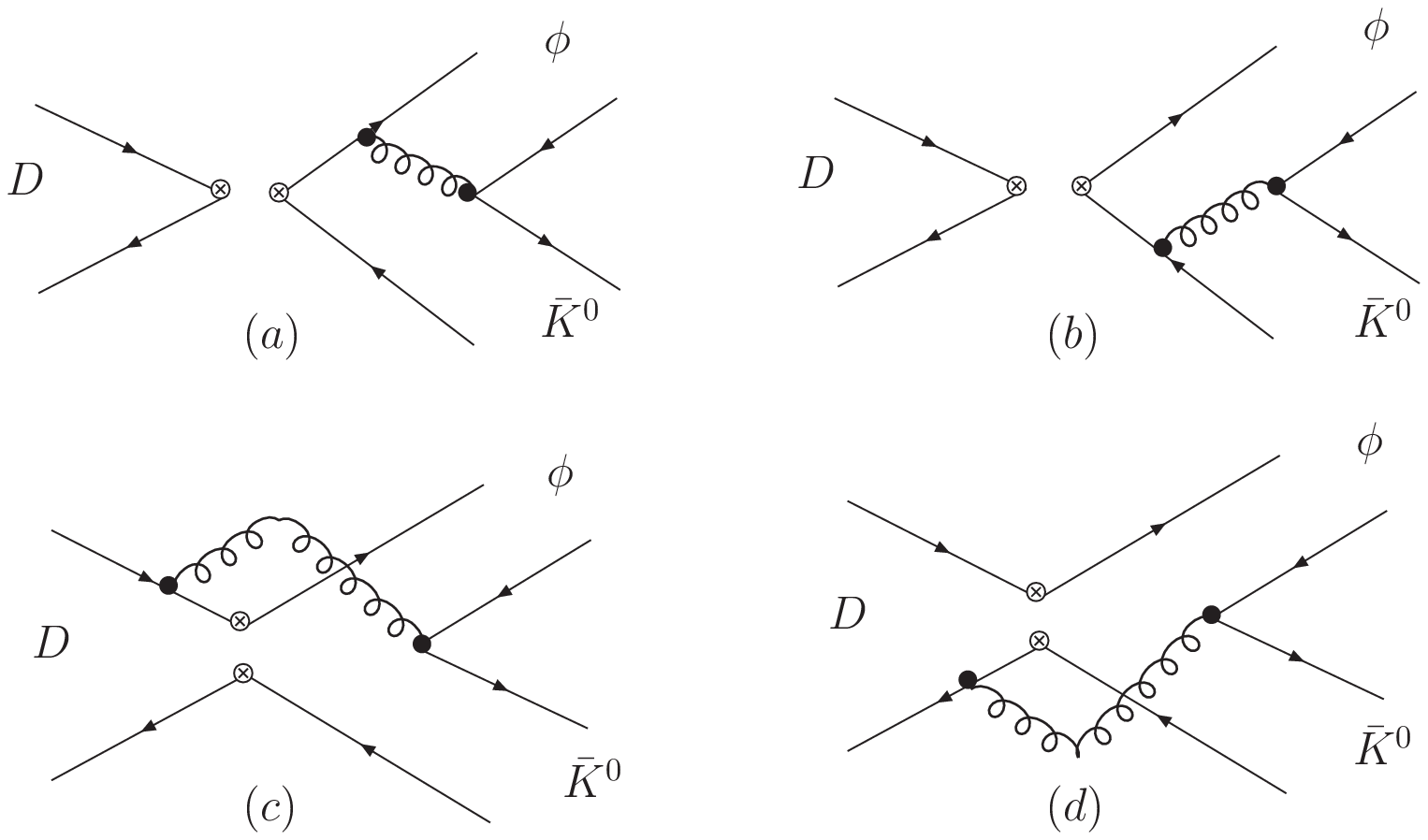}
\caption{Leading order Feynman diagrams of $D^0 \to \phi \overline K^0$}
\label{fig:diagrams1}
\end{center}
\end{figure}

For this decay, the relevant weak Hamiltonian is
\cite{Buchalla:1996vs}:
\begin{equation}
 H_\mathrm{eff} = \frac{G_F}{\sqrt{2}}V_{cs}V_{ud}^* \left[
C_1(\mu) O_1(\mu) + C_2(\mu) O_2(\mu) \right],\label{hami}
\end{equation}
 where $C_{i}(\mu)(i=1,2)$ is the QCD corrected Wilson coefficient at the
  renormalization scale $\mu$ and the four quark operators $O_{1}$ and $O_{2}$ are
\begin{eqnarray}
  O_1 = (\bar{s}_id_i)_{V-A}(\bar{u}_jc_j)_{V-A},
  O_2 = (\bar{s}_ic_i)_{V-A} (\bar{u}_jd_j)_{V-A}.
 \end{eqnarray}
The four lowest order Feynman diagrams of $D^0 \to \phi \overline
K^0$ in PQCD approach are drawn in FIG.\ref{fig:diagrams1} according
to this effective Hamiltonian.

By calculating the hard part $H$ at the first order of $\alpha_s$,
we get the following analytic formulas. With the meson wave
functions, the amplitude for the factorizable annihilation
diagrams in Fig.\ref{fig:diagrams1}(a) and (b) results in $F_a$
as:
\begin{widetext}
\begin{multline}
F_a= -16\pi C_F M_D^2 \int_0^1 \!\! dx_2 dx_3
 \int_0^\infty  \!\!  b_2 db_2\, b_3 db_3\times\Bigl[ E_{f}(t_a^1)
h_1(x_2,x_3,b_2,b_3)\bigl\{ (x_3-1)
  \phi_\phi(x_2)\phi_K^A(x_3)\\
  -2r
(x_3-2)r_K \phi_\phi^s(x_2)\phi_K^P(x_3)
-2x_3r r_K \phi_\phi^s(x_2)\phi_K^T(x_3)\bigr\}  -\bigl\{ (x_2-1)\phi_\phi(x_2)\phi_K^A(x_3) \\
 + 2x_2r
r_K\phi_\phi(x_2) \phi_K^P(x_3)
 - 2r(x_2-2) r_K\phi_\phi(x_2)
\phi_K^P(x_3) \bigr\} E_{f}(t_a^2) h_2(x_2,x_3,b_2,b_3) \Bigr].
\label{eq:Fa}
\end{multline}
\end{widetext}
In the function, $C_F = 4/3$ is the group factor of
$\mathrm{SU}(3)_c$, and $r_K = m_{0K}/M_D$, with $m_{0K} =
m_K^2/m_s$ ($m_s$ is the $s$ quark current mass). The functions
$E_{f}$ is :
\begin{equation}
 E_{f}(t) = (C_1(t)+ C_2(t)/3)\alpha_s(t)\ e^{-S_\phi(t)-S_K(t)}.
\end{equation}
The hard scale $t$'s in the amplitudes are taken as the largest
energy scale in the hard part $H$ in order to kill the large
logarithmic radiative corrections:
\begin{eqnarray}
 t_a^1 = \mathrm{max}(M_D \sqrt{1-x_3+x_3r^2},\frac{1}{b_2},\frac{1}{b_3}),
 t_a^2 = \mathrm{max}(M_D \sqrt{1-x_2},\frac{1}{b_2},\frac{1}{b_3}).
\end{eqnarray}
The functions $h_i(i=1,2)$  in the decay amplitudes consist of two
parts: one is the jet function $S_t(x_i)$ derived by the threshold
resummation, the other is the Fourier transformation of the propagator of virtual quark and gluon.
They are given as:
\begin{widetext}
\begin{align}
& h_1(x_2,x_3,b_2,b_3) = S_t(1-x_3)\left( \frac{\pi i}{2}\right)^2
H_0^{(1)}(M_D\sqrt{(1-x_2)(1-x_3+x_3r^2)}\, b_2) \nonumber \\
&\times \left\{ H_0^{(1)}(M_D\sqrt{1-x_3+x_3r^2}\, b_2)
J_0(M_D\sqrt{1-x_3+x_3r^2}\, b_3) \theta(b_2 - b_3) + (b_2
\leftrightarrow b_3 ) \right\}, \label{eq:propagator1}
\end{align}
\begin{align}
& h_2(x_2,x_3,b_2,b_3) = S_t(1-x_2)\left( \frac{\pi i}{2}\right)^2
H_0^{(1)}(M_D\sqrt{(1-x_2)(1-x_3+x_3r^2)}\, b_2) \nonumber \\
&\times \left\{ H_0^{(1)}(M_D\sqrt{1-x_2}\, b_2)
J_0(M_D\sqrt{1-x_2}\, b_3) \theta(b_2 - b_3) + (b_2
\leftrightarrow b_3 ) \right\}.
\end{align}
The amplitude for the nonfactorizable annihilation diagrams in
Fig.\ref{fig:diagrams1}(c) and (d) results in
\begin{multline}
M_a  =  \frac{1}{\sqrt{2N_c}} 64\pi C_F M_D^2 \int_0^1 \!\! dx_1
dx_2 dx_3
 \int_0^\infty \!\! b_1 db_1\, b_2 db_2\
\phi_D(x_1,b_1)  \\
\times \Bigl[ \bigl\{\left(x_3-1 \right)
\phi_\phi(x_2)\phi_K^A(x_3)
 - r \left(x_2+x_3-2 \right) r_K \phi_\phi^s(x_2)\phi_K^P(x_3)
 - r \left(x_2-x_3   \right) r_K \phi_\phi^t(x_2)\phi_K^P(x_3)\\
  - r \left(x_2+x_3-2 \right) r_K \phi_\phi^t(x_2)\phi_K^T(x_3)
 - r \left(x_2-x_3   \right) r_K \phi_\phi^s(x_2)\phi_K^T(x_3)
\bigr\}
E_{m}(t_{m}^1) h_a^{(1)}(x_1, x_2,x_3,b_1,b_2) \\
- \bigl\{\left(x_2-1 \right) \phi_\phi(x_2)\phi_K^A(x_3)
 - r \left(x_2+x_3-4 \right) r_K \phi_\phi^s(x_2)\phi_K^P(x_3)
 - r \left(-x_2+x_3   \right) r_K \phi_\phi^t(x_2)\phi_K^P(x_3)\\
  - r \left(x_2-x_3 \right) r_K \phi_\phi^s(x_2)\phi_K^T(x_3)
 + r \left(x_2+x_3   \right) r_K \phi_\phi^t(x_2)\phi_K^T(x_3)
\bigr\} E_{m}(t_{m}^2) h_a^{(2)}(x_1, x_2,x_3,b_1,b_2) \Bigr],
\label{eq:Ma1}
\end{multline}
where $x_1$ dependence in the numerators of the hard part are
neglected by the assumption $x_1 \ll x_2, x_3$. In the above
Equation, some functions are defined as:
\begin{eqnarray}
E_{m}(t)& =& C_2(t) \alpha_s(t)\, e^{-S_D(t)-S_\phi(t)-S_K(t)}; \\
t_{m}^j &= &\mathrm{max}(M_D \sqrt{|F^2_{(j)}|}, M_D
\sqrt{(1-x_2)(1-x_3+x_3r^2)}, \frac{1}{b_1},\frac{1}{b_2});
\end{eqnarray}
\begin{align}
& h^{(j)}_a(x_1,x_2,x_3,b_1,b_2) = \nonumber \\
& \biggl\{ \frac{\pi i}{2}
\mathrm{H}_0^{(1)}(M_D\sqrt{(1-x_2)(1-x_3+x_3r^2)}\, b_1)
 \mathrm{J}_0(M_D\sqrt{(1-x_2)(1-x_3+x_3r^2)}\, b_2) \theta(b_1-b_2)
\nonumber \\
& \qquad\qquad\qquad\qquad + (b_1 \leftrightarrow b_2) \biggr\}
 \times\left(
\begin{matrix}
 \mathrm{K}_0(M_D F_{(j)} b_1), & \text{for}\quad F^2_{(j)}>0 \\
 \frac{\pi i}{2} \mathrm{H}_0^{(1)}(M_D\sqrt{|F^2_{(j)}|}\ b_1), &
 \text{for}\quad F^2_{(j)}<0
\end{matrix}\right),
\label{eq:propagator2}
\end{align}
\end{widetext}
with:
\begin{eqnarray}
 F^2_{(1)}= (-x_1+x_2+1)(1-x_3+x_3r^2),
F^2_{(2)} =x_3(x_2-x_1)(1-r^2)-1.
\end{eqnarray}
The total decay amplitude for $D^0 \to \phi \overline K^0$ decay
is given as $ A = f_D F_a + M_a $. The decay width is then
\begin{equation}
 \Gamma(D^0 \to \phi \overline K^0) = \frac{G_F^2 M_D^3}{128\pi} (1-r^2)
|V_{ud}^*V_{cs} A|^2. \label{eq:neut_width}
\end{equation}

\begin{figure}[thb]
\begin{center}
\includegraphics[scale=0.55]{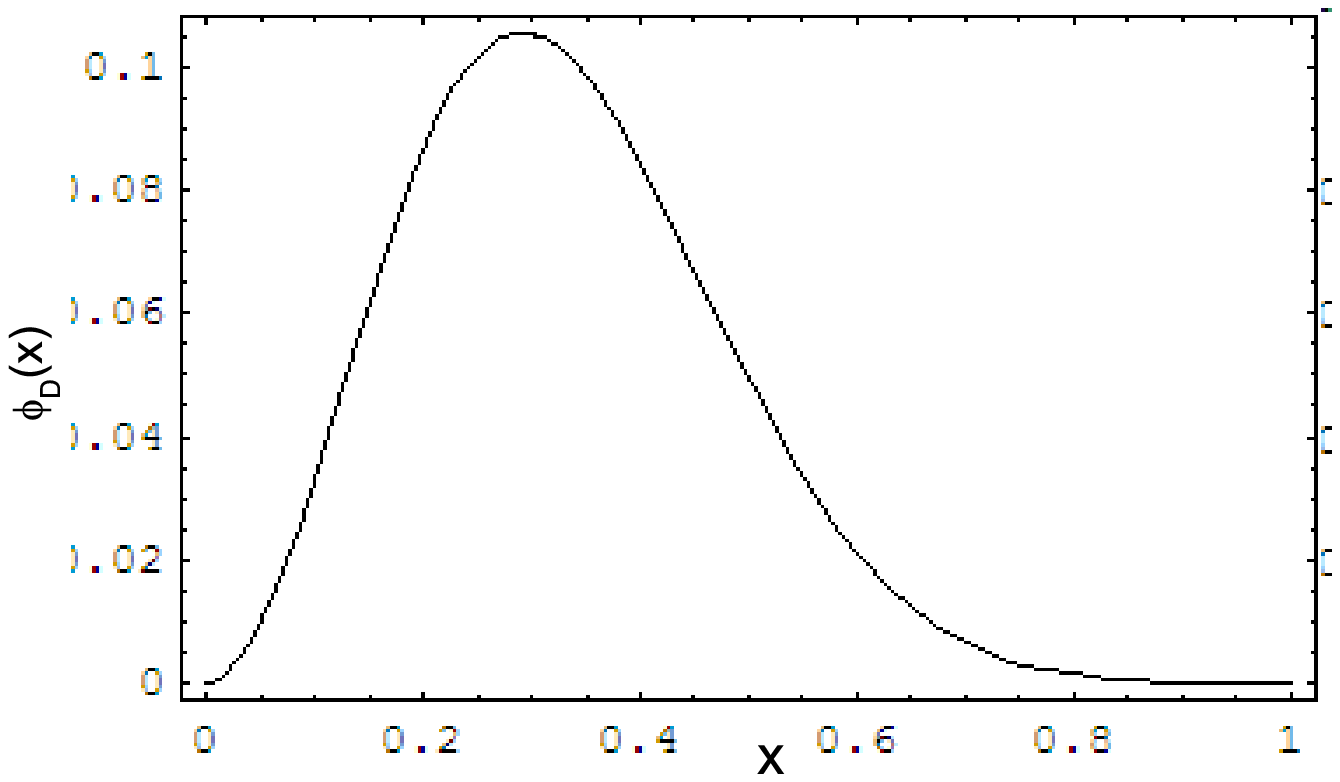}
\caption{The  $D$ meson  distribution amplitude} \label{fig2}
\end{center}
\end{figure}

For numerical analysis, we use the following input parameters
\cite{pdg}:
\begin{gather}
f_D = 230 \mbox{ MeV},~f_K = 160 \mbox{ MeV},~ f_{\phi} = 241 \mbox{
MeV},\nonumber
m_{0K} =1.7 \mbox{ GeV}, \nonumber\\ |V_{ud}|=0.9734, ~|V_{cs}|=0.974,
\tau_{D^0}=4.1\times 10^{-13}\mbox{s},~ f_{\phi}^T = 220 \mbox{MeV}.
\label{eq:shapewv}
\end{gather}
The branching ratio obtained from the analytic formulas may be
sensitive to many parameters especially those in the meson wave
functions. The light $\phi$ and $K$ mesons' distribution amplitudes
up to twist-3 have been used for many times in $B$ decays \cite{pqcd},
and we don't list them here. The  meson ($\phi$, $K$) wave functions
are well constrained by hadronic $B$ decays experiments.

 However, the heavy
meson wave functions are still in discussion especially for $D$
meson. In this work, the $D$ meson distribution amplitude we used
has only one parameter $\omega_D$, which is similar to $B$ meson
\cite{pqcd}:
\begin{equation}
\phi_D(x,b) = N x^2(1-x)^2 \exp \left[ -\frac{M_D^2\ x^2}{2
\omega_D^2} -\frac{1}{2} (\omega_D b)^2 \right],
\end{equation}
where $N$ is a normalization factor. For $B$ meson, the peak appears
at $x=0.1$, because $b$ quark is much heavier than the light $d$
quark. For $D$ meson, the ratio of heavy $(c)$ and light quark $(u)$
mass is rather smaller than that of $B$ meson. So we adjust the
parameter $\omega_D = 0.5 \mbox{ GeV}$, which makes the distribution amplitude
peak at $x=0.3$.
The shape of the $D$ meson distribution amplitude is shown in FIG.2.
Using this wave function, we can get the form factor of $D \to K $
as $F^{D\to K} =0.80$. This result is consistent with the previous
calculation $F^{D\to K} = 0.78\pm 0.04$ \cite{formf}. In fact, the
heavy meson distribution amplitude can be eventually determined by
radiative leptonic $D$ meson decays\cite{song}.

Here if we let $\omega_D$ vary
 from $0.45$ to $0.55$, the branching ratios of $D^0 \to \phi \overline K^0$ decay is:
\begin{gather}
\mathrm{Br}(D^0 \to \phi \overline K^0) = (8.7 \pm{1.6}) \times
10^{-3} ,
\end{gather}
which is consistent with the experimental measurement shown in
eq.(\ref{exp}). From this calculation, we find that the branching
ratio becomes large when $\omega_D$ arise. Many other parameters
such as Chiral breaking scale $m_{0K}$, CKM matrix elements also
have large uncertainties, which will also enlarge the theoretical
uncertainties for branching ratios.

In general, if a process happens in an energy scale where there are
many resonance states, this process must be seriously affected by
these resonances. This is a highly non-perturbative strong
interaction effect.  Near the scale of $D$ meson mass many resonance
states exist, which may give large pollution to $D$ decays
calculation.  So the final states interaction (FSI) may be
important. However, we cannot calculate the FSI's contribution from
the first principle. Although many people have discussed final
states interaction in $D$ decay \cite{FSI}, there is large
uncertainty in the calculation because they are usually model
dependent. In our PQCD calculation we factorize the non-perturbative
effects in meson wave functions, but neglect the soft FSI effect.
Our numerical result is consistent with the experimental data well.
It is a hint, that the contribution from soft FSI may  not play an
important role in this decay. We think the CLEO-C and BES can
measure this $W$-change channel more accurately, which will afford
help for us to understand the dynamics of $D$ decay. And the results
also help us determine the $D$ meson distribution amplitude. Since
there is only tree operators contributing to this decay (only one
kind of weak phase), there is no direct $CP$ violation in the standard
model. Any non-zero measurement of direct $CP$ in this decay will be a
signal of new physics.

In a summary, we calculate the branching ratio of $D^0 \to \phi
\overline K^0$ in the perturbative  QCD factorization approach
without considering final states interaction. Our result indicates
this branching ratio is very large comparing with other pure
annihilation decay channels, because there is no Cabibbo
suppression. This branching ratio is about $8.7 \times 10^{-3}$, and
has been measured by CLEO. We hope the CLEO-C and BES-III can
measure it more accurately, which will help us test QCD dynamics.

\end{document}